\documentclass[%
 reprint,%
 amssymb, amsmath,%
 aip,cha,%
]{revtex4-1}

\usepackage{bm}%
\usepackage[colorlinks=true,linkcolor=blue]{hyperref}%

\usepackage{graphicx}
\usepackage{amsmath}
\usepackage{subfigure}
\usepackage{color}
\usepackage{epsfig} 

\newcommand{\pvec}[1]{\vec{#1}\mkern2mu\vphantom{#1}}

\expandafter\ifx\csname package@font\endcsname\relax\else
 \expandafter\expandafter
 \expandafter\usepackage
 \expandafter\expandafter
 \expandafter{\csname package@font\endcsname}%
\fi
\begin{document}

\title{Ultrasoft classical systems at zero temperature.
}%

\author{Matheus de Mello}%
\affiliation{Departamento de F\'\i sica, Universidade Federal de Santa Catarina, 88040-900 Florian\'opolis, Brazil\looseness=-1}%

\author{Rogelio D\'iaz-M\'endez}%
\affiliation{Department of Physics, KTH Royal Institute of Technology,
	SE-10691 Stockholm, Sweden\looseness=-1}%

\author{Alejandro Mendoza-Coto}%
\email{alejandro.mendoza@ufsc.br}
\affiliation{Departamento de F\'\i sica, Universidade Federal de Santa Catarina, 88040-900 Florian\'opolis, Brazil\looseness=-1}%


\begin{abstract}
At low temperatures ultrasoft particle systems develop interesting phases via the self-assembly of particle clusters.
In this study we develop a general zero-temperature analysis fully characterizing the ground state of such models in two and three dimensions, considering the classical system with the restriction of a constant integer number of particles per cluster. 
We show that this methodology allows for an exact prediction of the actual density values at which the different phases emerge, including the zones of uneven cluster occupation, which are studied as coexistence regions of two pure phases.
Beyond the method itself, designed to produce exact phase diagrams from general ultrasoft potentials, we reach analytical expressions for the energy and location of the different phases in the large occupancy limit.
\end{abstract}

\maketitle


\section{Introduction}
Bounded repulsive interaction potentials whose Fourier transform has a negative minimum are known to be responsible for the self-assembly of cluster structures in systems of fully-penetrable particles.\cite{likos01,likos07,coslovich2012,prestipino14}
In the last decades, these so-called ultrasoft systems have emerged as a preferential model to study interesting collective phenomena that belong to apparently distant domains; from cold atoms,\cite{cinti14,dm15,pupillo20} vortex matter\cite{varney2013hierarchical,dm17,dm19w1pre} and nuclear matter\cite{horowitz04,caplan17,caplan18} to colloidal and polymeric systems.\cite{coslovich11,mladek08,dm19sm,zhuang17,lindquist17}
In this way the interest in describing the general properties of ultrasoft models and its connections to specific hamiltonians include both theoretical and practical motivations.

While the specific description of self-assembly processes is a general open subject in complex systems science, the motivation to measure cluster-forming experimental setups has raised up in the recent years within soft and condensed matter.\cite{likos16,colla18,clemens19}   
In all cases, the cluster formation that occurs at subcritical temperatures in ultrasoft systems closely resembles the properties and topological signatures of the corresponding ground-state phases.
It is then straightforward that a full understanding of the zero-temperature diagram of the cluster phases can boost the general comprehension of this emergent phenomenology.

In two and three dimensions, with increasing density, ultrasoft systems undergo an infinite sequence of transitions between cluster-crystalline states with increasing occupancy number.\cite{likos01b,likos07}
That is, between crystal structures in which the nodes consist of particle clusters of different size or occupancy, i.e.~number of particles per cluster.
Interestingly, these transitions occur without changing the underlying crystalline structure, but only increasing the cluster occupancy in a discontinuous, first-order character. 
The latter has been well-documented via phase diagrams obtained with computer simulations in a number of ultrasoft systems.\cite{zhang10,prestipino14,wilding14}
From these numerical studies it is clear that phases with integer occupation, i.e.~phases in which all clusters have exactly the same number of particles, emerge within intervals (instead of sharp values) of density.
However, this interplay between phases with integer and fractional occupation, which is key for the accurate description of the low-temperature phase diagrams, has been elusive for theoretical treatments to date.

In this work we present an exact analytical calculation to determine the zero-temperature phase diagram of any ultrasoft potential.
Since the clustering mechanism is produced to minimize the repulsion between particles, these systems arrange the clusters in a triangular lattice for two dimensions and a face-centered cubic (FCC) in three dimensions. 
By focusing on states of clusters with constant integer occupancy, referred to as pure phases, our formalism determine the energetic characteristics of these structures to explore its stability as a function of the density.
First order transitions from a cluster-crystal phase with occupancy $n$ to another of $n+1$ occur through a coexistence region that can be characterized by a fractional occupancy.
Therefore, using purely thermodynamical principles, a very general result can be obtained describing the emergence of both pure and mixed phases.

As a significant proof, the method is applied to the generalized exponential model GEM-$\alpha$, which is a well known cluster-forming interaction whose low-temperature phase diagram has been explored with numerical simulations for some values of $\alpha$.\cite{prestipino14}
We found a perfect agreement between our outcomes and the reported simulation results.
The additional finding of closed analytical expressions for the emergence of the different phases in the high-density regime also enlarge the relevance of the present study for the soft matter community.

\section{Analytical Description}
\label{sec2}

The interaction energy of a classical system of particles is given by
\begin{equation}
 E=\frac{1}{2}\sum_{i\neq j}V(\vec{r}_i-\vec{r}_j).
\end{equation}
We analyze separately the sequence of transitions occurring for the triangular lattice of clusters in two dimensions and for the FCC cluster lattice in three dimensions. In the following subsections we present a full characterization of the zero temperature properties of the system assuming that clusters are formed by superimposed particles.

\subsection{Triangular cluster-crystal in two dimensions}
The energy per particle of a two dimensional triangular lattice of $n$-particle clusters, with lattice spacing $a_n$, is given by
\begin{equation}
 \frac{E}{N}=\frac{1}{2}\left(n\sum_{p,q}V(\vert\vec{r}_{p,q}\vert)-V(0)\right),
 \label{int1}
\end{equation}
where $\vec{r}_{p,q}=a_n(p\vec{e}_1+q\vec{e}_2)$, $p$ and $q$ are integers, and the basis vectors representing the triangular lattice are taken as $\vec{e}_1=(1,0)$ and $\vec{e}_2=(-1/2,\sqrt{3}/2)$. 
Since the system is organized in a triangular lattice of clusters formed by $n$ superimpossed particles, with lattice parameter $a_n$, it is straightforward to conclude that the average density of the system will be
\begin{equation}
 \rho=\frac{2n}{\sqrt{3}a_n^2}.
\end{equation}
This relation allows to calculate the lattice parameter $a_n(\rho)$, at a given density, for configurations with any cluster occupation.  
The expression (\ref{int1}) can still be written in a more interesting way by rewriting it in terms of the Fourier transform of the interaction potential $\hat{V}(\vec{k})$, defined as 
\begin{equation}
 \hat{V}(k)=\int\frac{d^2k}{(2\pi)^2}e^{i\vec{k}\cdot\vec{r}}V(r).
\end{equation}

Now we can take advantage of the identity
\begin{equation}
 \sum_{p,q}V(\vert\vec{r}_{p,q}\vert)=\frac{2}{\sqrt{3}a_n^2}\sum_{p,q}\hat{V}(\vert\vec{k}_{p,q}\vert),
 \label{sumI}
\end{equation}
where the set of wave vectors $\vec{k}_{p,q}=k_0(n,\rho)(p\pvec{e}'_1 +q\pvec{e}'_2)$ define a triangular lattice in momentum space. 
Considering our previous choice of $\vec{r}_{p,q}$, the basis vectors of the set $\vec{k}_{p,q}$, can be taken as $\pvec{e}'_1=(0,1)$ and $\pvec{e}'_2=(\sqrt{3}/2,-1/2)$, and the lattice size is given by $k_0(n,\rho)={4\pi}/({\sqrt{3}a_n(\rho))}$. 
In appendix \ref{ap1} we present a demonstration of the identity of Eq.~(\ref{sumI}) which allows to rewrite the energy per particle of the triangular lattice as
\begin{equation}
 \frac{E}{N}=E_n=\frac{1}{2}\left(\rho\sum_{p,q}\hat{V}(\vert{\vec{k}_{p,q}}\vert)-V(0)\right).
\end{equation}
 The above expression corresponds to the exact energy of a pure triangular cluster lattice with a given integer cluster occupation number and density. As we will see, this information is sufficient to calculate all the ground-state properties including the classical ground-state phase diagram.
 The first building block to calculate the properties of the sequence of phase transitions, occurring as density is increased, will be the study of the crossing energies between energy curves corresponding to clusters with integer consecutive occupation numbers.
 In the next section we focus on the analytical treatment of this problem.    

\subsubsection{Energy crossing densities}
In general, each first-order transition occurring as the density is increased is given by a crossover region in which the occupancy number grows continuously from a given integer value $n$ to its consecutive $n+1$. 
A first estimation of the location of these phase transitions can be obtained calculating the densities at which the energy curves corresponding to consecutive integer occupation number crosses.  
Imposing the condition $E_n(\rho_n)=E_{n+1}(\rho_n)$ we reach to the equation
\begin{equation}
 f_V(k_0(n,\rho_n))=f_V(k_0(n+1,\rho_n)),
 \label{rec}
\end{equation}
where 
\begin{equation}
f_V(k)=\sum_{p,q}\hat{V}(k\vert{p\pvec{e}'_1 +q\pvec{e}'_2}\vert).
\end{equation}

Since $k_0(n,\rho)/k_0(n+1,\rho)=\sqrt{(n+1)/n}$, we know that in the limit $n\rightarrow \infty$, $k_0(\rho,n)\rightarrow k_0(n+1,\rho)$. 
This implies that, if Eq.~(\ref{rec}) have a sequence of solutions consistent with repeated transitions from clusters with  occupation $n$ to $n+1$, the function $f_V(k)$ must have a local maximum or minimum at some value $k_m$, around which the sequence of values of $k_0(n,\rho_n)$ and $k_0({n+1},\rho_n)$ at the transition, converge to $k_m$ as $n\rightarrow\infty$. 
Considering that in our case an increase of the density always produce transitions in which the cluster occupation number increase from $n$ to $n+1$, it can be concluded that $f_V(k)$ must have a local minimum at $k_m$. 
Consequently, around $k=k_m$, $f_V(k)$ can be approximated by a quadratic form of the type 
\begin{equation}
f_V(k)=f_V(k_m)+a/2(k-k_m)^2.
\end{equation}

From Eq.~(\ref{rec}) and considering that for large enough $n$ the values of $k_0(n,\rho_n)$ and $k_0(n+1,\rho_n)$ at the transition are close to $k_m$, we can reach to the condition $k_0(n,\rho_n)+k_0(n+1,\rho_n)=2k_m$. 
This condition allows to calculate the density $\rho_n$ at which the energies of the $n$ and $n+1$ phases coincide, which yields
\begin{equation}
 \rho_n=\frac{\sqrt{3}k_m^2n(n+1)}{2\pi^2\left(\sqrt{n+1}+\sqrt{n}\right)^2}.
\end{equation}
This result is expected to be valid in the large $n$ limit. Nevertheless, several cases were tested, producing relatively good estimates even for $n=1$. 

Regarding the dependence of the ground-state energy with the density, in general we already know that
\begin{equation}
 E_n(\rho)=\frac{1}{2}\left(\rho f_V(k_0(n,\rho))-V(0)\right),
 \label{ErhoG}
\end{equation}
 which is valid for densities $\rho$ in the interval $\left(\rho_{n-1},\rho_{n}\right)$. 
 In the asymptotic regime $n\gg1$, Eq.~(\ref{ErhoG}) can be approximated by
\begin{equation}
 E_n(\rho)=\frac{1}{2}\left(\rho (f_V(k_m)+\frac{a}{2}(k_0(n,\rho)-k_m)^2)-V(0)\right).
 \label{Erho}
\end{equation}
The family of energy functions defined by Eq.~(\ref{ErhoG}) described the ground state energy of the system considering only pure phases, i.e.~without considering the existence of coexistence regions. 

It is worth to compare these results with the classical mean-field results, in which $n$ is taken as a variational parameter. 
In the latter case it is straightforward to conclude that $n$ varies continuously with the density, in such a way that $k_0(n,\rho)=k_m$. 
This means that in the mean-field approximation
\begin{equation}
E_{MF}(\rho)=\frac{1}{2}\left(\rho f_V(k_m)-V(0)\right).
\end{equation}
Interestingly, this curve represents the envelope curve of the family of curves $E_n(\rho)$ given by Eq.~(\ref{Erho}). 
Moreover, at low enough densities we can see that $E_{MF}(\rho)<0$, which is clearly a drawback of the mean-field approach, since for purely repulsive potentials the ground state energy will be always a positive definite quantity. 

\subsubsection{Coexistence regions}
\label{coex}
As mentioned above, the first-order transitions between different cluster-crystal phases, as the density is increased, occur through a coexistence region in which the occupancy number has a crossover from $n$ to $n+1$. 
The densities corresponding to the beginning and the end of the coexistence regions can be determined by means of thermodynamic principles. 
Within the coexistence regions the pressure $P(\rho)$ and the chemical potential $\mu(\rho)$ of each pure cluster phase are equal and remain constant while the density is increased along the whole coexistence region. 
The mathematical condition determining the densities $\rho_{1n}$ and $\rho_{2n}$ at the beginning and the end of the coexistence region corresponding to the transition of cluster with occupancy number $n$ to $n+1$ is given by  
\begin{eqnarray}
\nonumber
 P_n(\rho_{1n})&=&P_{n+1}(\rho_{2n})\\
\mu_n(\rho_{1n})&=&\mu_{n+1}(\rho_{2n}).
\label{syst}
\end{eqnarray}
These pressures and chemical potentials can be calculated using the relations $P_n(\rho)=\rho^2\frac{\partial E_n(\rho)}{\partial \rho}$ and $\mu_n(\rho)=E_n(\rho)+\rho \frac{\partial E_n(\rho)}{\partial \rho}$. 

The system of Eqs.~(\ref{syst}) can not be solved in general for an arbitrary potential, even considering the large $n$ limit of the functions $E_n(\rho)$ given in Eq.~(\ref{Erho}). 
To proceed we look for a solution for $\rho_{1n}$ and $\rho_{2n}$ in the form of a large $n$ expansion of the type  
\begin{eqnarray}
\nonumber
 \rho_{1n}&=&a_0n+a_1+\frac{a_2}{n}+\ldots\\
 \rho_{2n}&=&b_0n+b_1+\frac{b_2}{n}+\ldots\ ,
 \label{co2d}
\end{eqnarray}
and solve the set of Eqs.~(\ref{syst}) order by order. 
The structure of the perturbative solution of this system of equations allows to conclude that, up to $O(\frac{1}{n})$, we must keep terms up to $O((k-k_m)^3)$ in $f_V(k)$. 
The expansion of $f_V(k)$ up to third order reads
\begin{equation}
 f_V(k)\approx f_0+\frac{f_2}{2!}(k-k_m)^2+\frac{f_3}{3!}(k-k_m)^3+...,
 \label{fexpand}
\end{equation}
where $f_n=f_V^{(n)}(k_m)$ represent the $n$-derivative of the function $f_V(k)$ evaluated in $k=k_m$. 
Solving perturbatively the system of Eqs.~(\ref{syst}) in powers of $n$ we obtain
\begin{eqnarray}
\nonumber
 a_0&=&b_0=\frac{\sqrt{3}k_m^2}{8\pi^2}\\ \nonumber
 a_1&=&\frac{\sqrt{3}k_m^2f_0}{2\pi^2(f_2k_m^2+8f_0)}\\ \nonumber
 b_1&=&\frac{\sqrt{3}k_m^2(f_2k_m^2+4f_0)}{8\pi^2(f_2k_m^2+8f_0)}\\ 
 a_2&=&b_2=-\frac{k_m^2f_0^2(9f_2+f_3k_m)(3f_2k_m^2+8f_0)}{2\sqrt{3}\pi^2f_2(f_2k_m^2+8f_0)^3}.
 \label{a012b1}
\end{eqnarray}
The obtained result for the coexistence boundaries of the first-order transition from cluster with occupancy $n$ to $n+1$, given in Eq.~(\ref{co2d}) and Eq.~(\ref{a012b1}) was compared with the numerical solutions of Eqs.~(\ref{syst}) for some specific models and the agreement, even for $n=1$, is surprisingly good. Details of this comparison can be found in section \ref{numteor}. 

Finally, we turn to the question of how to determine the energy of the ground state within the coexistence region. 
To calculate exactly the behavior of the energy in this regime we take advantage of the fact that, within this region, the pressure of the system remains constant and equal to the coexistence pressure $P_c$. 
Integrating the equation defining pressure in the system of Eqs.~(\ref{syst}), considering $P_c$ a constant, we obtain that within the coexistent region 
\begin{equation}
 E_n(\rho)=E_n(\rho_{1n})+P_c\left(\frac{1}{\rho_{1n}}-\frac{1}{\rho}\right).
 \label{Erhoc}
\end{equation}
This result is completely general and valid in the two dimensional as well as in the three dimensional case.
 
\subsection{FCC cluster-crystal in three dimensions}

The procedures followed in the previous section to study the ground-state of cluster-forming systems in two dimensions can be generalized to the three dimensional case without major difficulties. 
Numerical simulations as well as direct calculations allows to conclude that, among all possible three-dimensional crystals, the one minimizing the energy of the system under consideration is the FCC lattice. 
This is actually not surprising since the FCC is one of the closest packed structures in three dimensions.

As before, the energy of a cluster-crystal of occupation with $n$ particles per lattice site is given by
\begin{equation}
 \frac{E}{N}=\frac{1}{2}\left(n\sum_{p,q,s}V(\vert\vec{r}_{p,q,s}\vert)-V(0)\right),
\end{equation}
where $\vec{r}_{p,q,s}$ represent the position of the clusters in a FCC structure. 
We choose $\vec{r}_{p,q,s}=a_n(p\vec{v}_1+q\vec{v}_2+s\vec{v}_3$), where the basis vectors are taken as $\vec{v}_1=(0,1,1)1/2$, $\vec{v}_2=(1,0,1)1/2$, and $\vec{v}_3=(1,1,0)1/2$. 

For a FCC lattice of cluster, with $n$ particles per site, the average density is given by
\begin{equation}
 \rho=\frac{4n}{a_n^3},
 \label{cons3d}
\end{equation}
where $a_n$ represent the lattice spacing of the structure. 
This relation allows to calculate the lattice spacing in terms of the particle occupation of the clusters and the average density. 

Following the same method described in the two dimensional case we can rewrite the energy per particle of the system in the form
\begin{equation}
 \frac{E}{N}=E_n=\frac{1}{2}\left(\rho\sum_{p,q,s}\hat{V}(\vert{\vec{k}_{p,q,s}}\vert)-V(0)\right),
\end{equation}
where the wave vectors in the sum are taken as $\vec{k}_{p,q,s}=(2\pi\sqrt{3}/a_n)(p\pvec{v}'_1+q\pvec{v}'_2+s\pvec{v}'_3)$, and the basis vectors are given by $\pvec{v}'_1=(1,1,-1)1/\sqrt{3}$, $\pvec{v}'_2=(1,-1,1)1/\sqrt{3}$ and $\pvec{v}'_3=(-1,1,1)1/\sqrt{3}$.

Now we can find the densities at which the pure phases changes  stability. At these densities the condition $E_n(\rho)=E_{n+1}(\rho)$ implies that 
\begin{equation}
 g_V(k_0(n,\rho_n))=g_V(k_0(n+1,\rho_n)),
 \label{rec1}
\end{equation}
where $g_V(k)=\sum_{p,q,s}\hat{V}(k\vert{p\pvec{v}'_1+q\pvec{v}'_2+s\pvec{v}'_3}\vert)$ and $k_0(n,\rho_n)=2\pi\sqrt{3}/a_n(\rho)$, where $a_n(\rho)$  represent the lattice spacing of the FCC lattice related to the average particle density by Eq.~(\ref{cons3d}). 
In this case, as in the two dimensional case, when $n\gg1$, $a_n(\rho)/a_{n+1}(\rho)\rightarrow1$. 
This means that once again, in order to have a sequence of cluster transitions with increasing density, $g_V(k)$ needs to have a minimum at some finite value $k_m$. 
Consequently, for values of $k$ close enough to $k_m$ we can approximate $g_V(k)$ by its expansion up to second order  
\begin{equation}
 g_V(k)=g_V(k_m)+\frac{a}{2}(k-k_m)^2.
\end{equation}

The condition given in Eq.~(\ref{rec1}) leads again to the conclusion that $k_0(n,\rho_n)+k_0(n+1,\rho_n)=2k_m$. 
This equation allows to find the density at the cluster transition in the large cluster occupation limit considering only pure phases
\begin{equation}
 \rho_n=\frac{4 k_m^3n(1+n)}{3\sqrt{3}\pi^3 (n^{1/3} + (1 + n)^{1/3})^3}.
\end{equation}

Analogous to the two dimensional case, the system of Eqs.~(\ref{syst}) can now be solved for the FCC cluster crystal in order to find the densities limiting the coexistence region for each first-order transition.
Expanding $g_V(k)$ up to third order we obtain 
\begin{equation}
 g_V(k)\approx g_0+\frac{g_2}{2!}(k-k_m)^2+\frac{g_3}{3!}(k-k_m)^3+...,
\end{equation}
where 
$g_n=g_V^{(n)}(k_m)$ represent the $n$-derivative of the function $g_V(k)$ evaluated in $k=k_m$.

The system of Eqs.~(\ref{syst}) is solved now perturbatively in powers of $n^{-1}$ order by order, considering that the densities defining the boundaries of the coexistence region, for each first order transition, have the form given in Eq.~(\ref{co2d}). 
This solution process leads to the following coefficients for the three dimensional case 
\begin{eqnarray}\label{eq:coef}
\nonumber
 a_0&=&b_0=\frac{k_m^3}{6\sqrt{3}\pi^3}\\ \nonumber
 a_1&=&\frac{\sqrt{3} k_m^3 g_0}{2\pi^3(g_2k_m^2+18g_0)}\\
 b_1&=&\frac{k_m^3(g_2k_m^2+9g_0)}{6\sqrt{3}\pi^3(g_2k_m^2+18g_0)}\\\nonumber 
 a_2&=&b_2=-\frac{3\sqrt{3}k_m^3g_0^2(12g_2+g_3k_m)(g_2k_m^2+6g_0)}{4\pi^3g_2(g_2k_m^2+18g_0)^3}.
\end{eqnarray}

As in the two dimensional case the analytical solutions found here showed very good agreement with the exact numerical results even for low values of the cluster occupancy number $n$. 
This kind of analytical expression can be also useful to gain insights of the behavior of the cluster crystals at low temperatures.  

\section{The GEM-$\alpha$ case of study}

In order to test our analytical approach, the method developed in section~\ref{sec2} is used to characterize exactly the classical ground-state of the GEM-$\alpha$ model, which is given by a pairwise interaction of the form
\begin{equation}
    V(r)=\exp(-r^\alpha).
    \label{gema}
\end{equation} 
This effective potential have been commonly used in the literature to model systems of dendrimers, star-shaped polymers and general colloidal and polymeric system.~\cite{coslovich11,mladek08} 
The GEM-$\alpha$ model represents a bounded, repulsive  interaction whose Fourier  transform  has  a  negative  minimum at some wave vector for $\alpha>2$ in all dimensions. 
It is therefore an ultrasoft model that presents cluster phases with superimposed particles of the type studied in the present work.    

\subsection{Two dimensional system}
  
\begin{figure}[ht]
	\includegraphics[width=1\columnwidth]{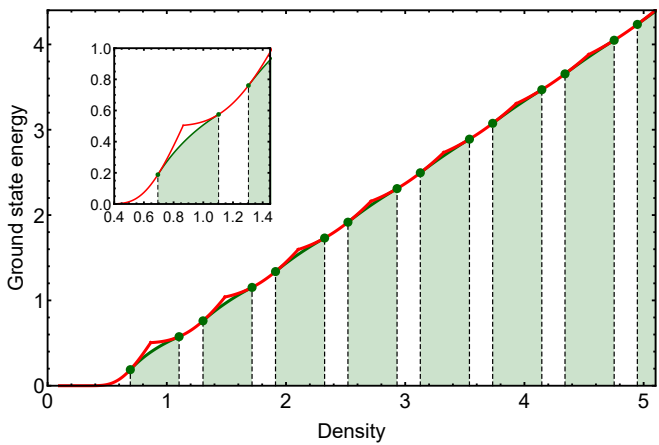}
    	\caption{
		Exact ground-state energy for the GEM-$4$ model. 
		The red curve corresponds to the ground-state energy of the system considering only pure cluster-crystal phases. 
		The change of stability for each pure cluster phase can be identified by the sharp peaks in the red curve. 
		The coexistence regions corresponding to each cluster crystal transition are represented by the shaded green areas. 
		In these regions, the actual ground-state energies are represented by the green solid curves. 
		The inset panel is a zoom of the original figure in a density interval corresponding to the energy branches $E_1(\rho)$ and $E_2(\rho)$, given by Eq.~(\ref{ErhoG}). 
		It is worth noting how the coexistence mechanism results in a further energy minimization when the energy of the mixed phase is compared with the energies of the pure phases involved in the phase transition. 
		}
		\label{fig1}
  \end{figure}

To illustrate the general behavior of the GEM-$\alpha$ model for a specific $\alpha>2$, in this section we study the exact properties of the GEM-4 model. 
In Fig.~\ref{fig1} we show the behavior of the ground-state energy in a wide range of densities. 
In this figure the red curve is formed by the different energy branches $E_n(\rho)$ corresponding to each pure cluster-crystal phase. 
The intersection between the different branches can be identified by the sharp peaks in the red curve (see the figure inset). 
At the same time, we understand that the first-order transition between the different cluster phases, as the density is increased, occurs through a coexistence region in which the cluster occupancy number varies from $n$ to $n+1$.  

The beginning ($\rho_{1n}$) and the end ($\rho_{2n}$) of each coexistence region can be determined numerically by solving the system of Eqs.~(\ref{syst}). 
Solving this system of equations allows to determine not only the boundaries of each coexistence region but also the pressure and the chemical potential within the coexistence region. 
Additionally, such information can be used to determine the behavior of the total energy per particle within the coexistence region by means of Eq.~(\ref{Erhoc}).

The coexistence region for each transition is represented by a shaded region in green. 
At the top of each shaded area, the exact value of the energy per particle is presented by the green solid curves. 
In the inset of Fig.~\ref{fig1}, we present a zoom of the original figure in a smaller density interval containing the first cluster transition, from the single particle triangular lattice to the two particle cluster-crystal. 
In this figure the differences between the energy curves corresponding to the pure phases and that of the coexistence region is evident. 
Additionally, the beginning and the end of each coexistence region is highlighted with green dots. 
As we can observe, the presence of the coexistence region results in a further minimization of the ground-state energy. 
From a thermodynamical point of view this is precisely why coexistence appears in a first-order transition: it is a mechanism of free energy minimization.

\begin{figure}[ht]
	\includegraphics[width=0.95\columnwidth]{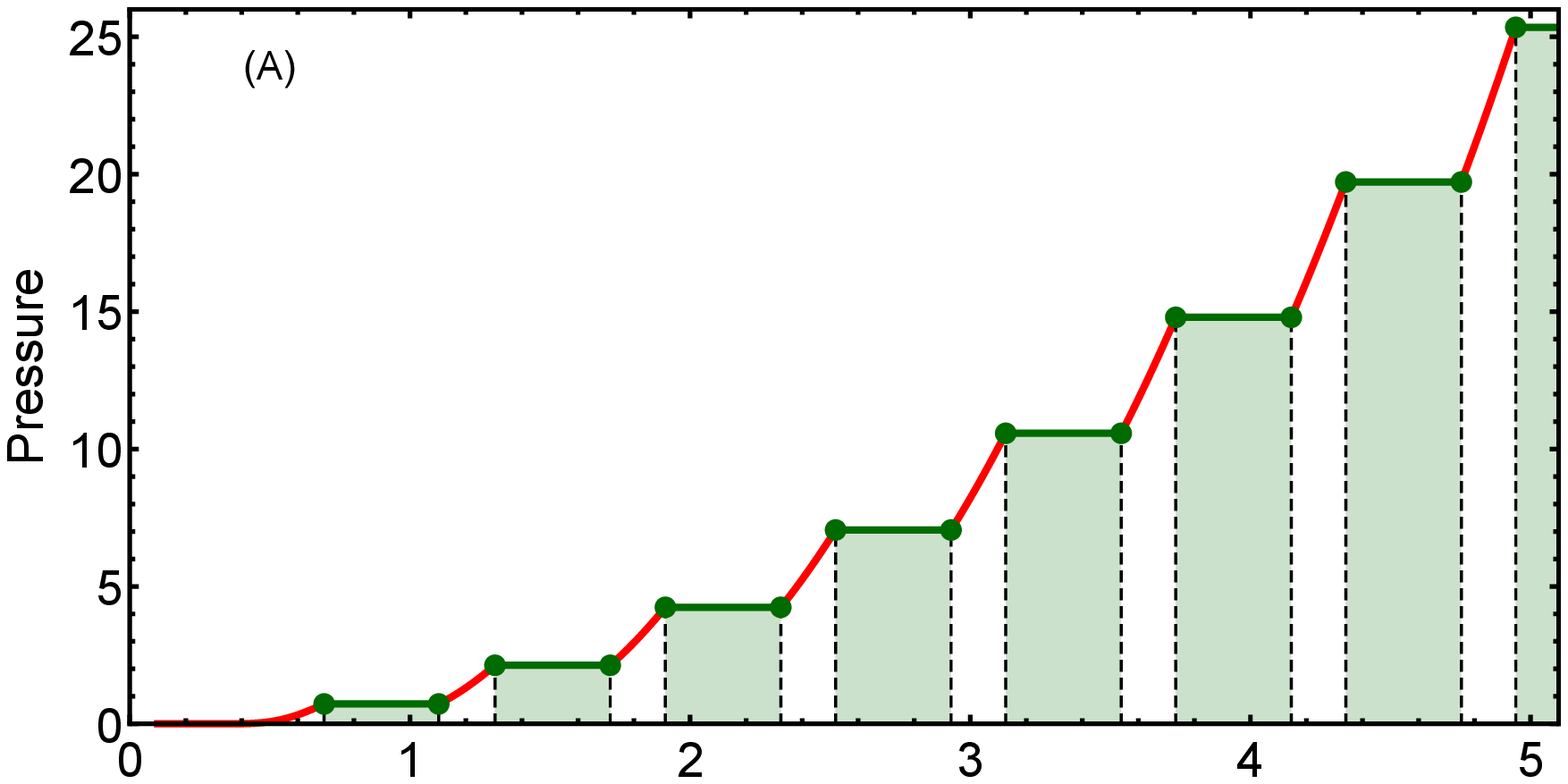}\vspace{0.3cm}
    \includegraphics[width=0.95\columnwidth]{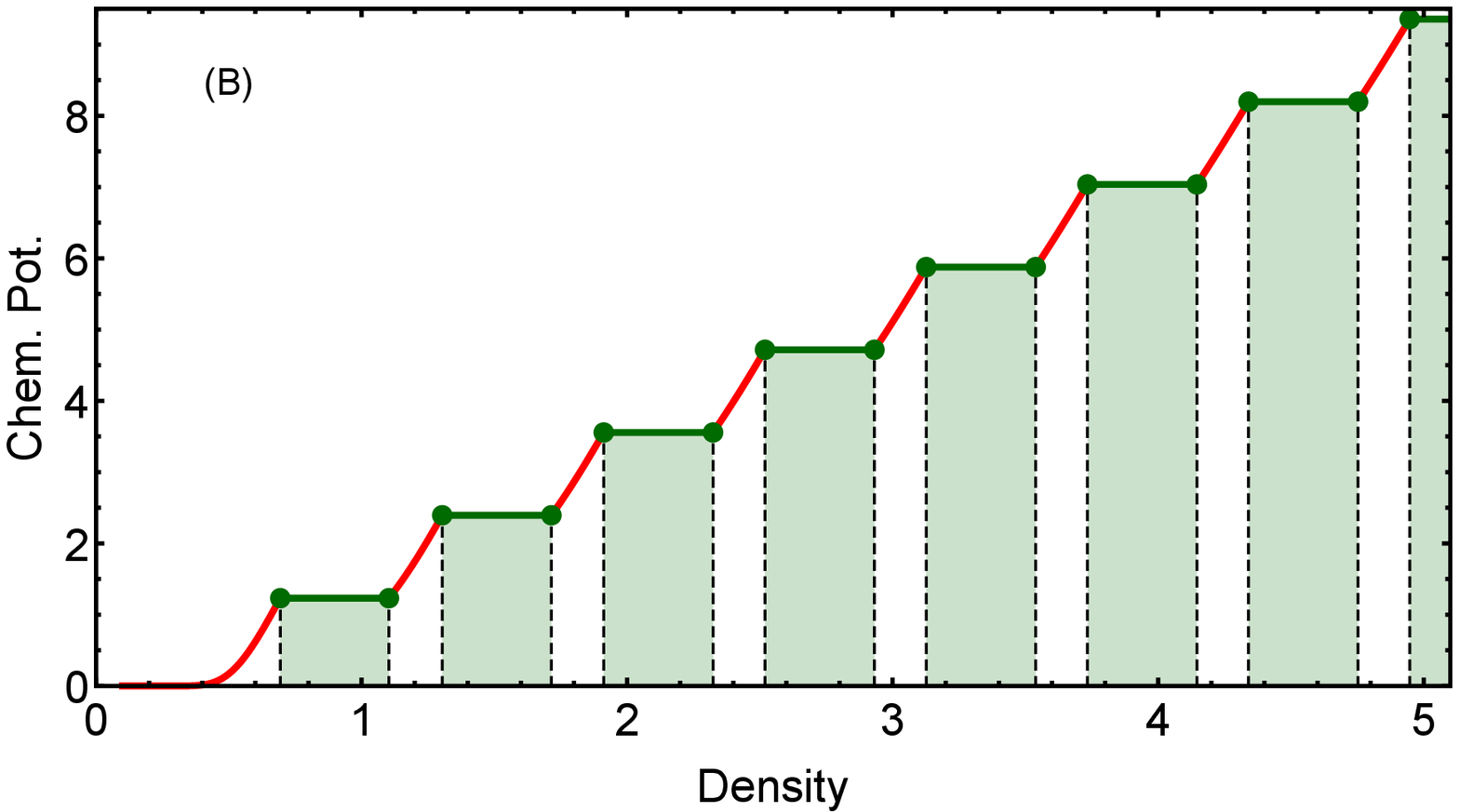}
	\caption{Pressure (A) and chemical potential (B) as a function of the density for the GEM-4 model in two dimensions. 
	Red curves in both figures correspond to the behavior of the specific magnitude whithin the pure cluster phases. 
	Green curves corresponds to the behavior of the specific magnitude within the coexistence regions. 
	The solid dots define the boundaries of the coexistence regions associated to each first order phase transition.}
	\label{fig:precoex}
\end{figure}

Once we have the exact ground-state energy curve we can calculate the corresponding pressure and chemical potentials by means of the definitions given in Eqs.~(\ref{syst}). In Fig.~\ref{fig:precoex}, we present the exact behavior of the pressure (panel A) and the chemical potential (panel B).
The red curves are associated to the behavior within pure cluster-crystal phases in which the occupancy number takes integer values. 
For states within the coexistence regions the curves of pressure and chemical potential remain constant as expected from thermodynamic principles. 
Once again the boundaries of the coexistence regions are highlighted with green points.

\begin{figure}[ht]
	\includegraphics[width=0.95\columnwidth]{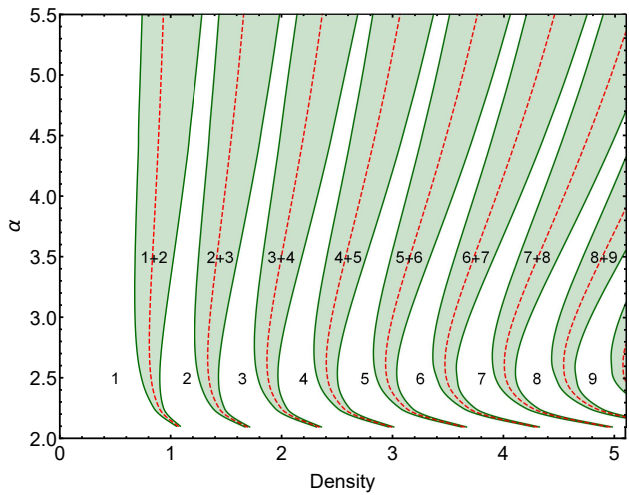}   
	\caption{Phase diagram $\alpha$ versus density for the GEM-$\alpha$ model in two dimensions. 
	For all $\alpha>2$ the system undergoes a infinite sequence of transitions between different cluster-crystal phases as the density is increased. 
	These transitions are accompanied by coexistence regions represented in the figure by the shaded green areas. 
	The dashed red curves represent the densities at which the pure cluster phases change stability.}
	\label{fig:phase2d}
\end{figure}

Now that the general properties of the ground state of the GEM-4 model have been described, we take one step further in the systematic characterization of the GEM-$\alpha$ model. 
In Fig.~\ref{fig:phase2d} we present the exact ground-state phase diagram varying the exponent $\alpha$ of the interaction potential and the particle density of the system. 
The shaded regions in green represent the coexistence regions associated to each first order transition between different cluster states. 
At the same time, the densities at which the pure phases changes stability, as the density is increased, is represented by red dashed lines. 
As expected, this lines are always found within the coexistence region corresponding to each first order phase transition. 

It can be noted that, as the value of $\alpha\rightarrow 2$, the position of the phase transitions moves progressively to infinity while the coexistence regions shrinks to zero. 
This behavior can be understood considering the properties 
of the function $f_V(k)$ in the limit $\alpha\rightarrow2$. 
In this regime, $k_m(\alpha)\rightarrow\infty$ and $f_2(k_m,\alpha)\rightarrow0$, this implies according to Eqs.~(\ref{a012b1}) that $a_0\rightarrow\infty$ and $a_1\rightarrow b_1$, which explain the observed behavior of the coexistence regions.

\subsubsection{Comparison between analytical and numerical results}
\label{numteor}

In this section we compare some of our analytical predictions with its numeral counterparts for the GEM-$\alpha$ model in the two-dimensional case. For the purpose of comparison we re-obtain the expression of the density $\rho_n$ at the crossing of the branches $E_n(\rho)$ and $E_{n+1}(\rho)$ in the two dimensional case. We look for a solution in the form of a large $n$ expansion of the same form of Eq.~(\ref{co2d})
\begin{equation}
 \rho_n=d_0n+d_1+\frac{d_2}{n}.
 \label{ranalytic}
\end{equation}
In this case the perturbative solution of Eq.~(\ref{rec}) order by order leads to the coefficients
\begin{eqnarray}
\nonumber
 d_0&=&\frac{\sqrt{3}k_m^2}{8\pi^2}\\ \nonumber
 d_1&=&\frac{\sqrt{3}k_m^2}{16\pi^2}\\
 d_2&=&-\frac{k_m^2(9\sqrt{3}f_2+\sqrt{3}k_m^2f_3)}{384\pi^2f_2},
 \label{eqe2}
\end{eqnarray}
where the coefficients $f_n$ and $k_m$ are defined in Eq.~(\ref{fexpand}).

\begin{figure}[ht]
	\includegraphics[width=1\columnwidth]{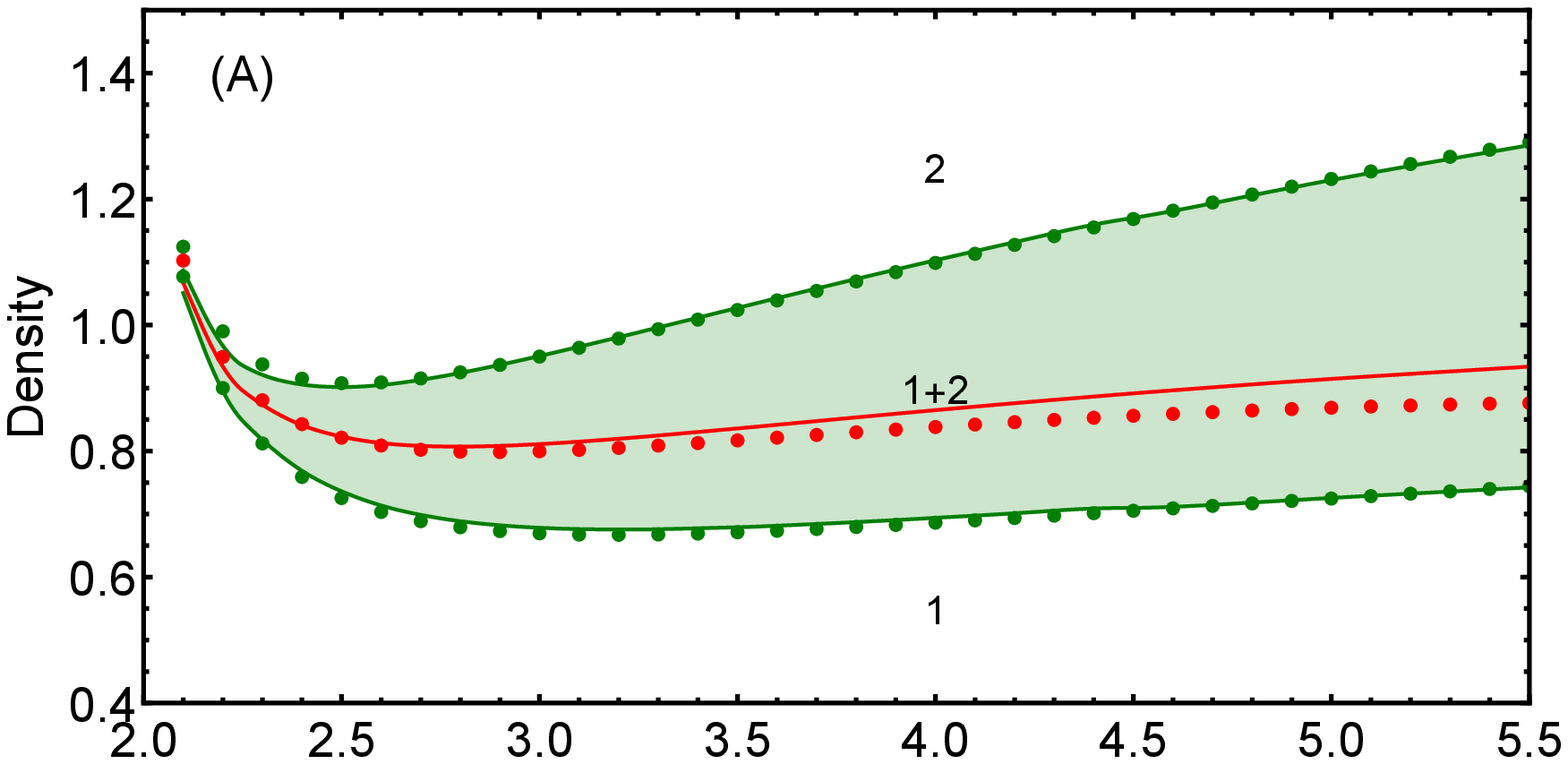}\vspace{0.3cm}
	\includegraphics[width=1\columnwidth]{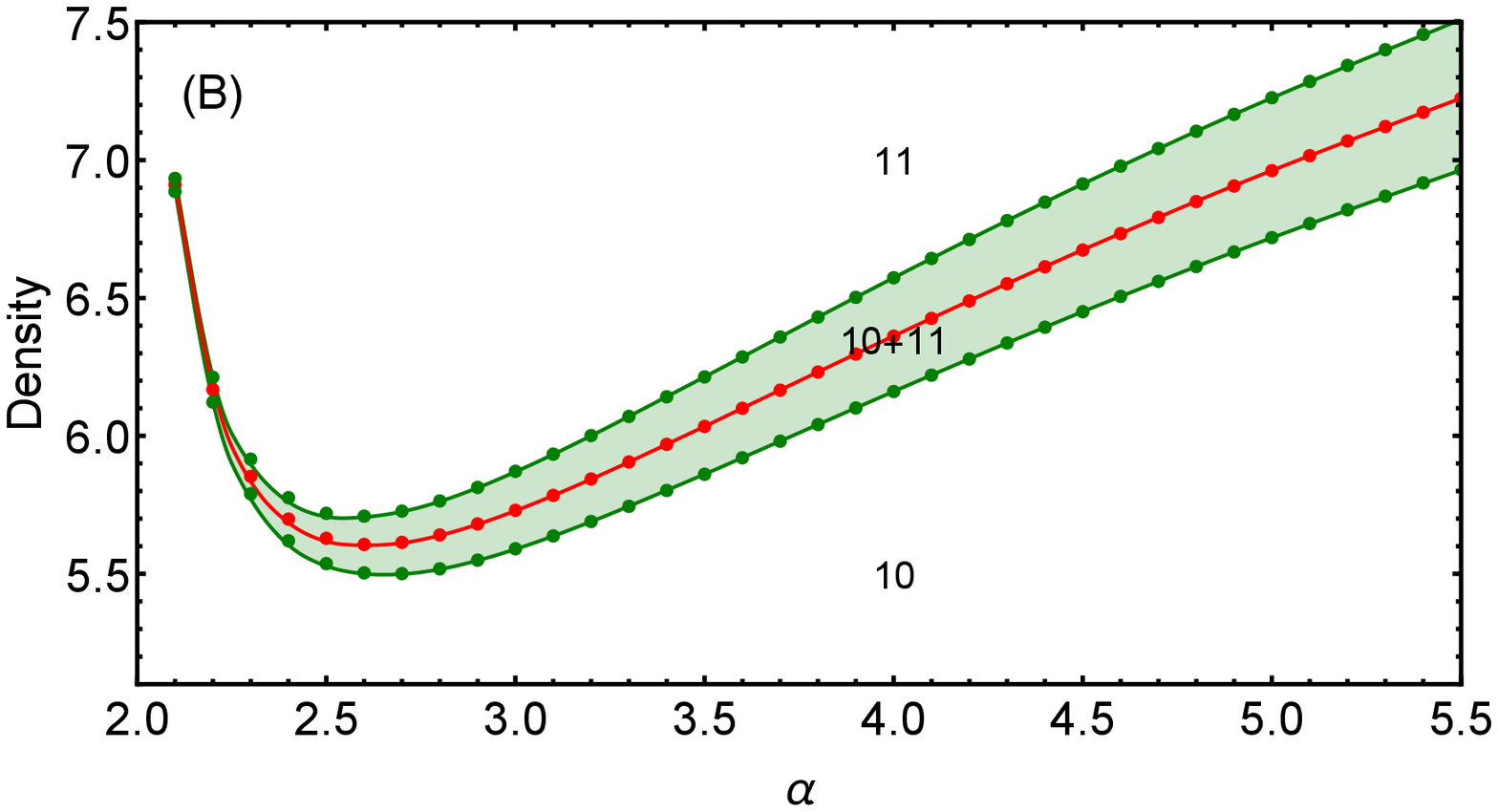}
    	\caption{Comparison of the analytical and exact numerical predictions for the densities as a function of the exponent $\alpha$ of the GEM-$\alpha$ model for specific transitions between different cluster states. 
    	The red dots represent the numerical exact values of the densities at the crossing of consecutive energy branches, while the continuous red curve represents the analytical asymptotic prediction given by Eqs.~(\ref{ranalytic},\ref{eqe2}). 
    	The green dots corresponds to the numerical exact values of the boundaries of the coexistence region while the green continuous lines are given by the analytic predictions in Eqs.~(\ref{co2d},\ref{a012b1}). 
    	Panel (A) corresponds to the transition from triangular lattice $n=1$ to the two particle cluster $n=2$. 
    	Panel (B) corresponds to the transition from the cluster crystal with ($n=10$) to the cluster-crystal with $n=11$.}
    	\label{fig4}
\end{figure}

In Fig.~\ref{fig4} we show a comparison between the analytical predictions and the exact numerical results for the GEM-$\alpha$ model in two dimensions. 
We show two different phase transitions: from the simple crystal $n=1$ to the two particle cluster-crystal $n=2$ (panel A), and the transition form the cluster-crystal $n=10$ to the cluster-crystal $n=11$ (panel B). 

The dotted curves represent the numerical results and the continuous curves represent the analytical large $n$ expressions given by Eqs.~(\ref{co2d},\ref{a012b1})  for the coexistence boundaries, and Eqs.~(\ref{ranalytic},\ref{eqe2}) for the densities at the energy crossing. 
Green full curves corresponds to the boundaries of the coexistence regions in each case, while red full curves are related to the energy crossing of the two relevant phases involved in the phase transition.

As can be observed, there is a high degree of coincidence between the numerical results and the analytical predictions obtained in the large $n$ limit already for the $n=10$ case, within the full range of models considered. 
On the other hand, for the lowest possible value of the cluster occupancy $(n=1)$, there is still a good agreement between analytical predictions and numerical exact results in the whole range of $\alpha$ considered, especially regarding the description of the boundaries of the coexistence region.        

\subsection{Three dimensional system}
 
 For completeness, in this section we apply the formalism described in Sec.~\ref{sec2} to study the GEM-$\alpha$ model in three dimensions.
 For values $\alpha\leq2$, the function $g_V(k)$ does not have a minimum at a finite value $k_m$, and therefore the system orders in a simple, non-cluster crystalline state. 
 Nevertheless, as density increases in this regime, a structural transition occurs from a FCC to a BCC structure accompanied by a very narrow coexistence region. 
 For $\alpha>2$, there is an infinite sequence of transitions between different FCC cluster-crystal phases in which the occupation number of the clusters increases with density. In this regime, the FCC cluster structure always have lower energy than the BCC cluster arrange. 
 
\begin{figure}[ht]
	\includegraphics[width=1\columnwidth]{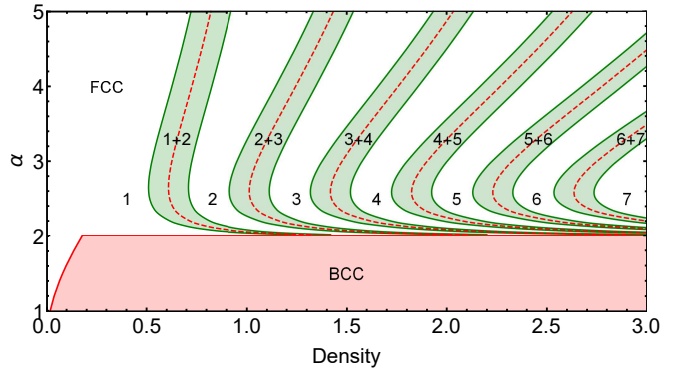}   	
		\caption{Phase diagram $\alpha$ versus density for the GEM-$\alpha$ model in three dimensions. 
		For values of $\alpha\leq2$ the system orders in a single particle array presenting a structural transition from a FCC lattice to a BCC lattice, for high enough densities. 
		For $\alpha>2$, a sequence of different FCC cluster crystals phases occurs as the density is increased. 
		These phases are represented by the white areas in the figure and have a well defined number of particles per clusters that increases with density. 
		The dashed red curves represent the densities at which the pure cluster phases changes stability, while the green shaded regions represent the coexistence regions associated to each cluster-crystal phase transition.}
		\label{fig:3d}
\end{figure}

 In Fig.~\ref{fig:3d} the $\alpha$ versus density phase diagram is constructed for the 3D system, comparing the energy per particle of the different cluster-crystal phases organized in a FCC lattice and in a single particle BCC lattice. 
 In the $\alpha\leq2$ regime, as mentioned before, there is no cluster formation and instead only a first order structural phase transition takes place from the FCC lattice to the BCC lattice as the density is increased. 
 The coexistence region associated to this transition is very narrow and consequently it is barely visible in the Fig.~\ref{fig:3d}.  
 
 In the regime $\alpha>2$ in which cluster formation occurs , the scenario is similar to the one observed in the two dimensional case. 
 In Fig.~\ref{fig:3d} the shaded green areas represent the different coexistence region associated to each cluster transition. 
 The red dashed lines, as before, provide the boundaries at which the different pure cluster crystal phases change stability. 
 Finally, the white regions of the phase diagram correspond to pure FCC cluster crystal phases.   
 
\section{Concluding remarks}

We have presented a study of the classical ground-state of particle systems interacting via an ultrasoft pair-wise potential in two and three dimensions. 
The method depicted for the construction of phase diagrams is general and produce exact results. 
Additionally, we have found closed analytical expressions in the large cluster limit for the ground-state energy, as well as for the location of the coexistence regions.

Comparing these results with previous works in the literature we observed that, up to our knowledge, the occupancy number of clusters have been always considered as a variational real parameter in the calculation of the ground-state energy of the system. 
This assumption is an attempt of capturing naturally the existence of uneven cluster occupation states corresponding to the coexistence regions of the phase diagram. 
It is important to remark that any mean-field approach in which the occupancy number of the cluster is described as a variational parameter will only produce approximate results, valid in the large density limit. 
Another important drawback of most mean-field calculations is related to the impossibility of properly describe the first order transitions occurring between different cluster phases as the density is increased. 
In this context, coexistence regions are not usually identified, since pure clusters phases does not exist in extended regions of the phase diagrams. 

In this work the assumption of stability of cluster phases with an homogeneous occupancy number allows to conclude the existence of first-order transitions and not crossovers between the different clusters phases. 
The properties of these first-order transitions are calculated from very general thermodynamic considerations and consequently our results show full consistency with what it is expected from a physical point of view. 
All the analytical results show excellent agreement with the exact numerical results presented and with previous numerical simulations results obtained for specific GEM-$\alpha$ models.

\appendix

\section{Lattice sum identity}
\label{ap1}
In order to prove the identity of Eq.~(\ref{sumI}) let us consider the sum
\begin{equation}
 S=\sum_{n,m}g\Big((n\vec{e_1}+m\vec{e}_2)a\Big),
\end{equation}
where $\vec{e}_1=(1,0)$ and $\vec{e}_2=(\cos{\theta},\sin\theta)$, represent the basis vectors of the lattice over witch the sum is performed. Our goal in this appendix is to obtain an equivalent expression for $S$ in terms of the Fourier transform of $g(\vec{r})$ assuming that such function exists.
Note that in our case this is always possible since $g(\vec{r})$ remains finite in all points of the summation lattice.

Considering the following definition of Fourier transforms
\begin{eqnarray}
 \hat{g}(\vec{k})&=&\int d^2r\ e^{-i\vec{k}\cdot\vec{r}}g(\vec{r})\\
 g(\vec{r})&=&\int\frac{d^2k}{(2\pi)^2}\ e^{i\vec{k}\cdot\vec{r}}\hat{g}(\vec{k}),
\end{eqnarray}
the original sum can be rewritten as
\begin{equation}
 S=\int\frac{d^2k}{(2\pi)^2}\sum_{n,m} e^{i\vec{k}\cdot(n\vec{e_1}+m\vec{e}_2)a}\hat{g}(\vec{k}).
 \label{fsum}
\end{equation}
And taking advantage of the Dirac comb identity 
\begin{equation}
\sum_n e^{inx}=(2\pi)\sum_n\delta(x-2\pi n),
\end{equation}
the summation in Eq.~(\ref{fsum}) can be performed to reach to the following expression for $S$
\begin{eqnarray}
\nonumber
 S&=&\int\frac{d^2k}{(2\pi)^2}\hat{g}(\vec{k})(2\pi)^2\sum_{n,m} \delta(\vec{k}\cdot\vec{e}_1a-2\pi n)\\
 &\times&\delta(\vec{k}\cdot\vec{e}_2a-2\pi m).
 \label{fsum1}
\end{eqnarray}
Now it can be directly integrated over momenta yielding
\begin{equation}
 S=\frac{1}{a^2\vert\sin(\theta)\vert}\sum_{n,m}\hat{g}(\vec{k}_{n,m}),
\end{equation}
where
\begin{equation}
\vec{k}_{n,m}=\frac{2\pi}{a\vert\sin(\theta)\vert}(n\pvec{e}'_1+n\pvec{e}'_2),
\end{equation}
with  $\pvec{e}'_1=(0,1)$ and  $\pvec{e}'_2=(\sin(\theta),-\cos(\theta))$. 

It is now straight forward to check the validity of Eq.~(\ref{sumI}) just by considering that our sum is performed over a triangular lattice for which $\theta=2\pi/3$.

\section*{Author's Contributions}
All authors contribute equally to the presented work.

\section*{Data Availability}
The data that support the findings of this study are available from the corresponding author upon reasonable request.

\section*{Acknowledgements}

A.M.C. acknowledges financial support from Funda\c{c}\~ao de Amparo \`a Pesquisa de Santa Catarina, Brazil (FAPESC).
R.D.M. acknowledges the Swedish Research Council Grants No.~642-2013-7837,  2016-06122,  2018-03659 and G\"oran Gustafsson Foundation for Research in Natural Sciences and Medicine and Olle Engkvists Stiftelse.

\end{document}